\documentclass[sigconf]{acmart}

\usepackage{courier}  
\usepackage{url}  
\usepackage{graphicx} 
\usepackage{natbib}  
\usepackage{caption} 
\usepackage{algorithm}
\usepackage{algorithmic}
\usepackage{graphicx}   
\usepackage{tabularx}   
\usepackage{array}      
\usepackage{subcaption} 
\usepackage{comment}
\usepackage{multirow}
\usepackage{enumitem}
\usepackage{comment}

\usepackage{tikz}
\usetikzlibrary{positioning, arrows.meta, calc}

\newcolumntype{Y}{>{\raggedright\arraybackslash}X} 

\usepackage{newfloat}
\usepackage{listings}
\DeclareCaptionStyle{ruled}{labelfont=normalfont,labelsep=colon,strut=off} 
\floatstyle{ruled}
\newfloat{listing}{tb}{lst}{}
\floatname{listing}{Listing}
\setcopyright{acmlicensed}
\acmConference[FedKDD/FedMAS 2026]{Joint Workshop on Federated Learning for Multi-agent Systems and Data Mining}{August 9, 2026}{Jeju, Korea}

\begin{document}

\title{STAIN-FL: Stealthy Targeted Attack Injection with Contextual Triggers in Federated Learning}

\author{Ashlinder Kaur}
\affiliation{%
  \institution{Singapore Institute of Technology (SIT)}
  \country{Singapore}}
\email{2202636@sit.singaporetech.edu.sg}

\author{Purnima Murali Mohan}
\affiliation{%
  \institution{Singapore Institute of Technology (SIT)}
  \country{Singapore}}
\email{purnima.mohan@singaporetech.edu.sg}

\author{Zengxiang Li}
\authornote{Corresponding author.}
\affiliation{%
  \institution{SingHealth Duke-NUS AI in Medicine Institute}
  \institution{SingHealth AI Office, Singapore}
  \country{}
  }
\email{shawn.li.zx@singhealth.com.sg}

\author{Tram Truong-Huu}
\affiliation{%
  \institution{Singapore Institute of Technology (SIT)}
  \country{Singapore}}
\email{truonghuu.tram@singaporetech.edu.sg}

\begin{abstract}
Federated video anomaly detection trains model collaboratively without sharing raw surveillance footage, but limited server-side visibility lets compromised clients to inject backdoor via malicious updates. This paper introduces STAIN-FL, a stealthy targeted backdoor attack injection framework that uses naturally occurring surveillance conditions, including low-light scenes, indoor settings, and crowd density, as contextual triggers. STAIN-FL combines anomaly-to-benign label \textit{manipulation} with gradient masking over least-updated coordinates to preserve clean accuracy while inducing trigger-conditioned misclassification. We evaluate STAIN-FL on \texttt{UCF-Crime} using 1024-dimensional I3D features in a non-IID four-client multi-agency setting, comparing FedAvg and FedProx under sparse and continuous attacks. Results show that sparse attacks have low-detectability, operationally significant attacks rather than high-intensity attacks: they keep the mean clean-accuracy drop below $2\%$, yet still misclassify more than half of triggered anomalies at peak backdoor accuracy under FedAvg ($56.7\%$) and FedProx ($54.2\%$). Under FedAvg, the sparse backdoor remains above the $25\%$ backdoor-accuracy threshold for an average of $336$ post-attack rounds, highlighting the persistence risk of contextually triggered attacks in surveillance systems.

\end{abstract}

\maketitle
\section{Introduction}

AI-driven video surveillance has emerged as critical public safety infrastructure worldwide, with the global AI in video surveillance market projected to grow from USD $6.26$ billion in 2025 to USD $18.33$ billion by 2032 \cite{fortune_ai_surveillance_2025}. 
While AI-driven surveillance applications can support operations in multi-agency and cross-jurisdictional deployments, privacy and regulatory restrictions often limit centralization of sensitive surveillance data. 
This creates a practical need for learning frameworks that can exploit distributed surveillance data while preserving local data ownership and operational autonomy. Federated learning (FL) offers a privacy-preserving alternative by enabling collaborative model training across distributed edge devices and IoT endpoints without requiring raw data to leave each client.
However, conventional FL systems, such as FedAvg~\cite{mcmahan2023communicationefficientlearningdeepnetworks} and FedProx~\cite{li2020federatedoptimizationheterogeneousnetworks}, also introduce a structural backdoor vulnerability because the server observes only client updates, not the underlying data or local training process. This limited visibility can allow a compromised client to inject adversarial updates into the global model while evading direct inspection. 
In video surveillance settings, this risk is particularly concerning because an attacker may seek to suppress the detection of specific adverse conditions (e.g., low-light scenes, indoor settings, crowd density), and use these conditions as \textit{contextual triggers} to execute the attacks while preserving normal model performance on clean inputs.

Recent literature has increasingly examined backdoor vulnerabilities in FL~\cite{NGUYEN2024107166, 2024-40487}, with much of the work shaped by two challenges: how aggregation algorithms perform under heterogeneous data, and how injected backdoors can be made to persist beyond the attack phase. In terms of aggregation, FedAvg~\cite{mcmahan2023communicationefficientlearningdeepnetworks} and FedProx~\cite{li2020federatedoptimizationheterogeneousnetworks} remain widely adopted algorithms, with the latter introducing a proximal term to improve stability under non-IID conditions~\cite{article}. In this work, we focus on the latter and study attack methodologies. Neurotoxin~\cite{pmlr-v162-zhang22w} and Stealthy and Long-Lasting Durable Backdoor Attack in Federated Learning (SDBA)~\cite{Choe_2025} have shown that backdoors can be made durable by targeting gradient coordinates that are rarely updated by honest clients. SDBA further demonstrates that stealth and durability can be jointly achieved through layered gradient masking. However, these advances have been demonstrated primarily on NLP and image classification benchmarks.

Several limitations of the current literature motivate the present work. First, durable FL backdoor attacks have been studied mainly on natural language processing and image classification benchmarks~\cite{pmlr-v162-zhang22w,Choe_2025}. Their effectiveness on video-based anomaly detection, which is central to public safety surveillance, remains largely unexamined. Second, prior attacks typically use artificial triggers, such as pixel patterns or token insertions. These triggers may be easier to detect than contextually grounded triggers that exploit naturally occurring properties of surveillance videos, such as low-light scenes, indoor settings, and crowd density. Third, while existing studies have shown that backdoors can persist beyond the attack phase, the post-attack behavior of the global model is still not well characterized. In particular, it remains unclear whether the backdoor gradually decays, disappears, or stabilizes at a non-trivial equilibrium. These gaps are significant in multi-agency surveillance deployments, where a persistent and undetected backdoor in the shared global model could allow selected anomaly classes to evade detection across participating agencies.

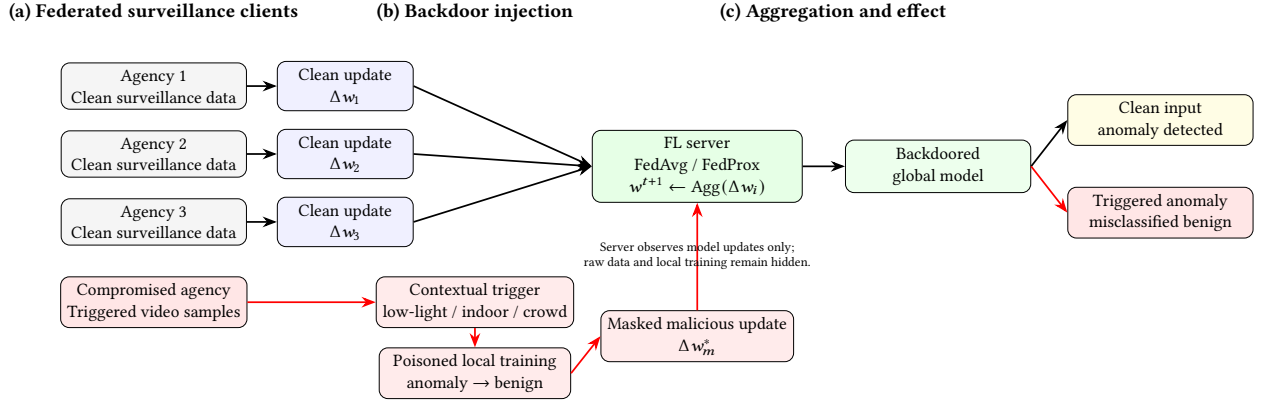
\begin{figure*}[t]
\centering
\resizebox{0.93\textwidth}{!}{%
\begin{tikzpicture}[
    font=\small,
    >=Stealth,
    client/.style={
        rectangle, draw, rounded corners,
        minimum width=3.0cm, minimum height=0.75cm,
        align=center, fill=gray!8
    },
    badclient/.style={
        rectangle, draw, rounded corners,
        minimum width=3.0cm, minimum height=0.75cm,
        align=center, fill=red!10
    },
    update/.style={
        rectangle, draw, rounded corners,
        minimum width=2.2cm, minimum height=0.65cm,
        align=center, fill=blue!6
    },
    attack/.style={
        rectangle, draw, rounded corners,
        minimum width=3.1cm, minimum height=0.75cm,
        align=center, fill=red!8
    },
    server/.style={
        rectangle, draw, rounded corners,
        minimum width=3.4cm, minimum height=1.0cm,
        align=center, fill=green!10
    },
    model/.style={
        rectangle, draw, rounded corners,
        minimum width=3.0cm, minimum height=0.85cm,
        align=center, fill=green!7
    },
    outcome/.style={
        rectangle, draw, rounded corners,
        minimum width=3.0cm, minimum height=0.75cm,
        align=center, fill=yellow!12
    },
    arrow/.style={->, thick},
    redarrow/.style={->, thick, red},
    group/.style={draw, rounded corners, dashed, inner sep=8pt}
]

\node[font=\bfseries] at (0,4.3) {(a) Federated surveillance clients};
\node[font=\bfseries] at (5.2,4.3) {(b) Backdoor injection};
\node[font=\bfseries] at (11.0,4.3) {(c) Aggregation and effect};

\node[client] (c1) at (0,3.1) {Agency 1\\Clean surveillance data};
\node[client] (c2) at (0,2.0) {Agency 2\\Clean surveillance data};
\node[client] (c3) at (0,0.9) {Agency 3\\Clean surveillance data};
\node[badclient] (cm) at (0,-0.4) {Compromised agency\\Triggered video samples};

\node[update] (u1) at (3.1,3.1) {Clean update\\$\Delta w_1$};
\node[update] (u2) at (3.1,2.0) {Clean update\\$\Delta w_2$};
\node[update] (u3) at (3.1,0.9) {Clean update\\$\Delta w_3$};

\draw[arrow] (c1.east) -- (u1.west);
\draw[arrow] (c2.east) -- (u2.west);
\draw[arrow] (c3.east) -- (u3.west);

\node[attack] (trig) at (5.2,-0.4)
{Contextual trigger\\low-light / indoor / crowd};

\node[attack] (poison) at (5.2,-1.55)
{Poisoned local training\\anomaly $\rightarrow$ benign};

\node[attack] (mask) at (8.8,-0.95)
{Masked malicious update\\$\Delta w_m^{*}$};

\draw[redarrow] (cm.east) -- (trig.west);
\draw[redarrow] (trig.south) -- (poison.north);
\draw[redarrow] (poison.east) -- (mask.west);

\node[server] (server) at (8.8,1.8)
{FL server\\FedAvg / FedProx\\
$w^{t+1} \leftarrow \mathrm{Agg}(\Delta w_i)$};

\node[model] (global) at (12.7,1.8)
{Backdoored\\global model};

\node[outcome] (clean) at (16.3,2.55)
{Clean input\\anomaly detected};

\node[outcome, fill=red!10] (bd) at (16.3,1.05)
{Triggered anomaly\\misclassified benign};

\draw[arrow] (u1.east) -- (server.west);
\draw[arrow] (u2.east) -- (server.west);
\draw[arrow] (u3.east) -- (server.west);
\draw[redarrow] (mask.north) -- (server.south);

\draw[arrow] (server.east) -- (global.west);
\draw[arrow] (global.east) -- (clean.west);
\draw[redarrow] (global.east) -- (bd.west);

\node[align=center, font=\scriptsize] at (8.8,0.35)
{Server observes model updates only;\\
raw data and local training remain hidden.};

\end{tikzpicture}%
}
\caption{STAIN-FL backdoor injection pathway in federated video anomaly detection.}
\label{fig:stain_fl_detailed}
\end{figure*}

Figure~\ref{fig:stain_fl_detailed} illustrates the STAIN-FL attack pathway considered in this work. A compromised client first constructs contextually triggered samples by associating naturally occurring surveillance conditions, such as low-light scenes, indoor settings, or crowd density, with a target benign label. The client then performs poisoned local training and submits a masked malicious update together with clean updates from honest clients. Since the FL server aggregates model updates without direct visibility into local data or the local training processes, the poisoned update may be incorporated into the global model. As a result, the backdoor model can preserve normal anomaly-detection performance while misclassifying trigger-conditioned anomalies as benign.
This paper addresses these gaps by introducing STAIN-FL, a stealthy targeted backdoor attack injection framework for federated video anomaly detection. The key innovation is to move beyond artificial trigger patterns and study contextually grounded triggers derived from naturally occurring surveillance conditions, namely low-light scenes, indoor settings, and crowd density. Using these adverse conditions to trigger attacks helps STAIN-FL maintain its persistence in attacks without being detected due to the low quality of the input data, which also causes the model performance degradation. STAIN-FL combines these triggers with gradient-masking principles inspired by durable FL backdoor attacks to inject malicious updates that preserve clean-model utility while inducing targeted misclassification. We evaluate the attack under both FedAvg and FedProx in a realistic non-IID multi-agency setting, and systematically characterize its post-attack behavior to determine whether the injected backdoor decays, disappears, or stabilizes after the attack phase ends. Through this framework, we investigate two research questions:

\begin{description}[leftmargin=0pt,itemsep=1pt,topsep=2pt]
    \item[\textbf{RQ1 (Stealth vs.\ Success):}] Can a backdoor attack achieve high backdoor accuracy while remaining undetected, with minimal degradation to the global model's main-task accuracy?
    \item[\textbf{RQ2 (Persistence):}] How long does a backdoor survive after the attack phase ends, and does it decay towards elimination or stabilise at a non-trivial equilibrium?
\end{description}

\section{Related Work}

Video anomaly detection (VAD) aims to identify abnormal events or objects in surveillance videos, such as assaults, robberies, accidents, and other public-safety incidents. 
Recent work has begun to study VAD under federated learning, motivated by the difficulty of centralizing sensitive video data across organizations. CLAP formulates collaborative anomaly learning with privacy for unsupervised VAD, while FedVAD and recent federated weakly supervised VAD methods address heterogeneous client data and privacy-preserving surveillance learning~\cite{allahham2024clap, wang2025federatedvad}. These studies show that FL is a promising architecture for distributed surveillance analytics, but they mainly focus on detection performance and generalization. The security behavior of federated VAD under targeted backdoor attacks remains comparatively underexplored.

FL is vulnerable to backdoor attacks because the server aggregates client updates without directly observing the data or the training process that produced them. A compromised client can therefore submit poisoned updates that preserve clean-task performance while causing attacker-desired behavior under a trigger condition. Existing studies and surveys show that FL backdoor attacks can arise through data poisoning, model poisoning, or hybrid strategies, and that non-IID client data makes malicious updates harder to distinguish from legitimate distributional variation~\cite{NGUYEN2024107166,li2025backdoor}. Recent KDD work has also examined context-dependent backdoor behavior in graph prompt learning, showing that backdoor risks extend beyond conventional image or text classification settings~\cite{lyu2024crosscontext}. However, most attacks are still evaluated on image, text, or graph benchmarks, rather than surveillance-oriented VAD.

A key challenge in FL backdoor attacks is durability: once the attacker stops injecting poisoned updates, subsequent benign training may overwrite the malicious behavior due to the forgetting issue of deep learning models. Neurotoxin addresses this by targeting parameters that are rarely modified by honest clients, thereby increasing the likelihood that the backdoor persists across later aggregation rounds~\cite{pmlr-v162-zhang22w}. SDBA extends this direction by combining layer-wise and top-$k$ gradient masking to improve both stealth and long-term durability \cite{Choe_2025}. These works establish that backdoor persistence depends not only on attack strength but also on where the malicious update is injected in the parameter space. STAIN-FL builds on this insight by studying whether durable gradient-masked backdoors can persist in federated video anomaly detection using naturally occurring surveillance conditions as contextual triggers.

\section{Threat Model and STAIN-FL Attack Procedure}
\begin{table}[t]
\centering
\caption{Notation used in the threat model}
\label{tab:symbols}
\vspace{-1.5ex}
\begin{tabular}{p{0.23\columnwidth}p{0.67\columnwidth}}
\toprule
\textbf{Symbol} & \textbf{Description} \\
\midrule
$K$ & Number of FL clients \\
$t$ & FL communication round \\
$m$ & Compromised client index \\
$\mathbf{w}^{(t)}$ & Global model at round $t$ \\
$\mathbf{w}_{m}^{(t)}$ & Local model trained by client $m$ \\
$\mathcal{D}_m$ & Dataset of compromised client \\
$\mathcal{D}_{clean}$ & Clean local samples \\
$\mathcal{D}_{bd}$ & Poisoned backdoor samples \\
$\mathcal{D}_{atk}$ & Clean and poisoned attack dataset \\
$\mathcal{C}$ & Candidate contextual triggers \\
$\tau$ & Selected contextual trigger \\
$\tau(x)$ & Trigger indicator for video $x$ \\
$k$ & Gradient mask ratio \\
$s$ & Malicious update scale factor \\
$\Delta \mathbf{w}_{m}^{(t)}$ & Poisoned update before masking \\
$\mathbf{M}^{(t)}$ & Least-updated-coordinate mask \\
$\Delta \mathbf{w}_{m}^{*(t)}$ & Final masked malicious update \\
$f(x)$ & Model prediction for video $x$ \\
$y$ & Label: $0$ benign, $1$ anomaly \\
\bottomrule
\end{tabular}
\vspace{-2.5ex}
\end{table}

We consider a cross-silo FL setting for video anomaly detection with $K$ agency clients. At communication round $t$, the server broadcasts the current global model $\mathbf{w}^{(t)}$ to all clients. Each client trains locally on its private surveillance dataset and returns a model update to the server. The server aggregates these updates using FedAvg or FedProx. The server is assumed to be honest, but it observes only model updates and has no direct access to client data, ground-truth labels, or local training processes.

\paragraph{\bf Adversary assumptions.}
We consider a single threat model and let $m$ denote the compromised client and $\mathcal{D}_m$ denote its local dataset. The adversary controls only client $m$. It can modify local ground-truth labels, change the local training process (e.g., modifying training parameters, such as gradient mask ratio, learning rate boost, described in Table~
\ref{tab:fixedparameters}), and submit a malicious update to the server. It cannot control the server, the aggregation rule, or the honest clients. The attacker knows its own dataset $\mathcal{D}_m$, the model architecture, and the FL training configuration, but does not require access to the private datasets of honest clients.

\paragraph{\bf Contextual trigger space.}
STAIN-FL defines a set of candidate contextual triggers
\[
\mathcal{C}=\{\texttt{low-light}, \texttt{indoor}, \texttt{crowded},~\dots~\}.
\]
Unlike artificial pixel patches or visible overlays, these triggers are naturally occurring surveillance conditions. The compromised client selects an active trigger $\tau \in \mathcal{C}$ for the attack. 

\paragraph{\bf Attack objective.}
The attack is targeted and aims to make the global model classify trigger-conditioned anomalous videos as benign while preserving normal behavior on clean inputs. For an anomalous video $x$ with label $y=1$, the adversary seeks
\[
f(x)=0 \quad \text{if} \quad \tau(x)=1.
\]
Thus, the backdoor should activate only when the selected contextual trigger is present.

\paragraph{\bf Poisoned local data.}
During an attack round, the compromised client constructs a poisoned subset from $\mathcal{D}_m$ by relabeling only the triggered anomalous videos as benign:
\[
\mathcal{D}_{bd}=\{(x,0)\mid (x,1)\in \mathcal{D}_m,\tau(x)=1\}.
\]
The local attack dataset is then
\[
\mathcal{D}_{atk}=\mathcal{D}_{clean}\cup\mathcal{D}_{bd},
\]
where $\mathcal{D}_{clean}$ contains the remaining clean samples. The compromised client trains locally on $\mathcal{D}_{atk}$ starting from $\mathbf{w}^{(t)}$.

\paragraph{\bf Masked malicious update.}
After local training, the compromised client obtains a poisoned update $\Delta \mathbf{w}_{m}^{(t)}$. To improve stealth and durability, STAIN-FL constructs a binary mask $\mathbf{M}^{(t)}$ over the least-updated $k\%$ of model coordinates. The final malicious update submitted to the server is
\[
\Delta \mathbf{w}_{m}^{*(t)}
=
s\left(\mathbf{M}^{(t)}\odot \Delta \mathbf{w}_{m}^{(t)}\right),
\]
where $s$ is the scale factor and $\odot$ denotes element-wise multiplication. This masking step is intended to place the poisoned signal in coordinates that honest clients are less likely to overwrite in later rounds.
Algorithm~\ref{alg:stain_fl} shows how the compromised client selects the contextual trigger, forms $\mathcal{D}_{bd}$ and $\mathcal{D}_{atk}$, computes the poisoned update, applies the least-updated-coordinate mask, and submits $\Delta \mathbf{w}_{m}^{*(t)}$ for FedAvg/FedProx aggregation.

\setlength{\textfloatsep}{5pt}
\begin{algorithm}[t]
\caption{STAIN-FL Backdoor Injection}
\label{alg:stain_fl}
\begin{algorithmic}[1]
\REQUIRE $\mathbf{w}^{(t)}, \mathcal{D}_m, \mathcal{C}, k, s$
\ENSURE $\Delta \mathbf{w}_{m}^{*(t)}$

\STATE $\tau \leftarrow \textsc{SelectTrigger}(\mathcal{C})$
\STATE $\mathcal{D}_{bd} \leftarrow \{(x,0)\mid (x,1)\in\mathcal{D}_m,\tau(x)=1\}$
\STATE $\mathcal{D}_{atk} \leftarrow \mathcal{D}_{clean}\cup\mathcal{D}_{bd}$
\STATE $\mathbf{w}_{m}^{(t)} \leftarrow \textsc{LocalTrain}(\mathbf{w}^{(t)},\mathcal{D}_{atk})$
\STATE $\Delta \mathbf{w}_{m}^{(t)} \leftarrow \mathbf{w}^{(t)}-\mathbf{w}_{m}^{(t)}$
\STATE $\mathbf{M}^{(t)} \leftarrow \textsc{LeastUpdatedMask}(k)$
\STATE $\Delta \mathbf{w}_{m}^{*(t)} \leftarrow s(\mathbf{M}^{(t)}\odot\Delta \mathbf{w}_{m}^{(t)})$
\STATE \textbf{return} $\Delta \mathbf{w}_{m}^{*(t)}$
\end{algorithmic}
\end{algorithm}

The server aggregates $\Delta \mathbf{w}_{m}^{*(t)}$ together with honest client updates without knowing that the update was trained on poisoned data. The resulting global model may therefore preserve clean-input accuracy while misclassifying trigger-conditioned anomalies as benign. We evaluate this behavior using clean accuracy, backdoor accuracy, and post-attack persistence.

\section{Experimental Evaluation}

\subsection{Experimental Setup}

We evaluate STAIN-FL using Flower for the FL training framework and PyTorch for local training and gradient computation. The system comprises four heterogeneous clients, each representing an independent surveillance data owner with a distinct local data distribution. This design captures the non-IID nature of multi-organization deployments while avoiding raw-data sharing. All clients participate in every communication round and perform one local epoch before returning model updates to the server.
We use the \texttt{UCF-Crime} dataset~\cite{ucfcrime}, comprising $1900$ real-world surveillance videos balanced between $950$ normal and $950$ anomalous samples spanning $13$ anomaly classes. Each video is converted into a fixed 1024-dimensional feature vector using a pre-trained Inflated 3D ConvNet (I3D), capturing the spatio-temporal content required for activity recognition. To emulate cross-silo heterogeneity, anomaly classes are partitioned across the four clients according to the incident types each client would naturally encounter, with a small number of classes shared to reflect overlapping client boundaries. 
A balanced global test set comprising $20\%$ of the data is held out (at server), while the remaining $80\%$ is distributed across clients and split locally into $80\%$ training and $20\%$ testing. Full per-client distributions are provided in Appendix~\ref{app:data}.
The attack is evaluated under both FedAvg and FedProx aggregation. 
We model a data-rich compromised client (instantiated as Client~1 in the experiments) to reflect the risk posed by a high-contribution participant. The compromised client uses the low-light trigger. We compare sparse and continuous attacks, vary the gradient mask ratio across five values, and evaluate 20 configurations in total. Each configuration is repeated five times and averaged; the experimental parameters are summarized in Table~\ref{tab:fixedparameters}. The implementation and experimental configurations are available at \url{https://github.com/Ashlinder/STAIN-FL}.

\begin{table}[h]
\centering
\caption{Experimental parameters}
\label{tab:fixedparameters}
\begin{tabular}{ll}
\toprule
Parameter & Value \\
\midrule
Compromised client & Agency~1 \\
Trigger type & \texttt{low-light} \\
Learning rate boost ($\alpha$) & 2.0 \\
Scale factor ($s$) & 1.5 \\
Attack start round & 50 \\
Total FL rounds & 1{,}000 \\
Gradient mask ratios ($k$) & 0.05, 0.10, 0.15, 0.20, 0.25 \\
Runs per configuration & 5 \\
\bottomrule
\end{tabular}
\vspace{-1.5ex}
\end{table}

\subsection{Evaluation Metrics}

We assess attack outcomes using clean accuracy, backdoor accuracy (BA), stealth, and durability. Clean accuracy is the global model's classification accuracy on the global test set, while BA is the proportion of triggered anomalous videos misclassified as normal.

\begin{itemize}[leftmargin=*, itemsep=1pt, topsep=2pt, parsep=0pt, partopsep=0pt]
    \item \textbf{Stealth:} measured as the average clean-accuracy drop during the attack phase relative to a five-round pre-attack baseline. An attack is considered stealthy if the drop remains below $5\%$, chosen to exceed the natural $\pm$2--3\% fluctuation observed under non-IID training (Appendix~\ref{app:threshold}).

    \item \textbf{Threshold-based durability:} measured as the number of post-attack rounds required for BA to fall and remain below selected thresholds: $50\%$, $40\%$, $30\%$, $25\%$, and $20\%$. This captures when the backdoor ceases to remain effective.

    \item \textbf{Volatility-based durability:} measured as the number of post-attack rounds required for the 10-round rolling standard deviation of BA to fall below selected tolerances: $\pm$10\%, $\pm$5\%, and $\pm$3\%. This captures when backdoor behavior stabilizes, regardless of the absolute BA level.
\end{itemize}

\subsection{Results and Analysis}
The results show that STAIN-FL’s main risk is not maximum attack intensity, but the combination of low detectability and long post-attack persistence under contextual triggers.

\begin{figure}[t]
\centering
\includegraphics[width=0.95\columnwidth]{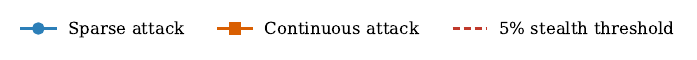}\\[-2pt]
\begin{subfigure}{0.495\columnwidth}
  \includegraphics[width=\textwidth]{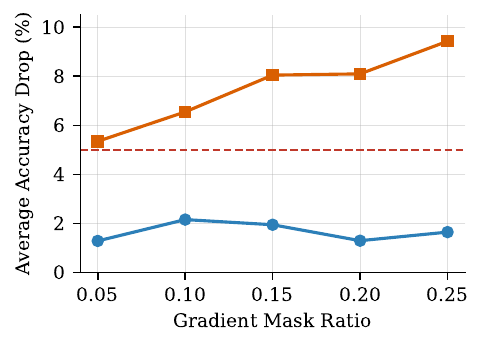}
  \caption{FedAvg}
  \label{fig:drop-fedavg}
\end{subfigure}
\hfill
\begin{subfigure}{0.495\columnwidth}
  \includegraphics[width=\textwidth]{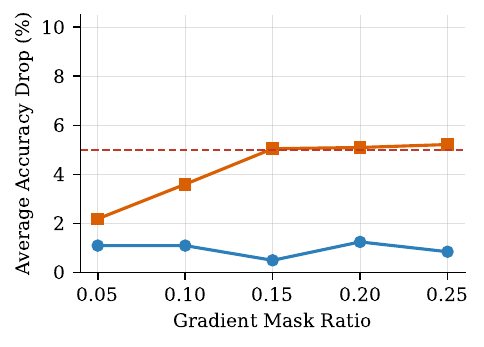}
  \caption{FedProx}
  \label{fig:drop-fedprox}
\end{subfigure}
\caption{Average accuracy drop across gradient mask ratios. The dashed line marks the 5\% stealth threshold; sparse attacks remain below it in all configurations.}
\label{fig:accdrop}
\end{figure}

\paragraph{\bf Stealth.}
Sparse attacks remain stealthy across all configurations and both aggregation algorithms, with mean clean accuracy drops of 1.66\% under FedAvg and 1.03\% under FedProx --- well below the 5\% threshold. Continuous attacks, on the contrary, are substantially more detectable: under FedAvg, the mean accuracy drop is 7.51\% and exceeds the threshold at every gradient mask ratio, rising steadily as the ratio increases (Figure~\ref{fig:accdrop}). FedProx reduces this mean drop to 4.23\% by constraining the magnitude of poisoned updates through its proximal term, bringing continuous attacks below the threshold only at low mask ratios, while larger ratios remain detectable. The intermittent nature of the sparse pattern provides stealth structurally: sparse attacks remain below the detection threshold regardless of aggregation algorithm or mask ratio.

\begin{figure}[t]
\centering
\includegraphics[width=0.75\columnwidth]{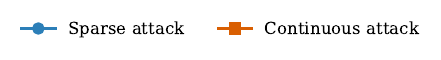}\\[-2pt]
\begin{subfigure}{0.495\columnwidth}
  \includegraphics[width=\textwidth]{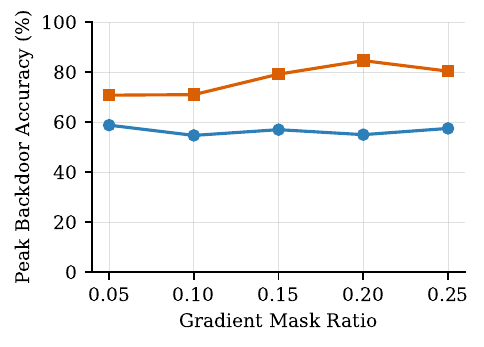}
  \caption{FedAvg}
  \label{fig:peak-fedavg}
\end{subfigure}
\hfill
\begin{subfigure}{0.495\columnwidth}
  \includegraphics[width=\textwidth]{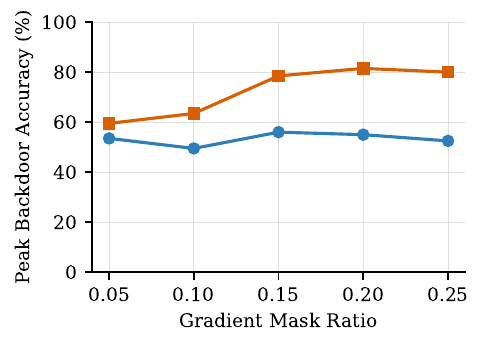}
  \caption{FedProx}
  \label{fig:peak-fedprox}
\end{subfigure}
\caption{Peak backdoor accuracy across gradient mask ratios. Continuous attacks achieve substantially higher peak effectiveness than sparse attacks under both algorithms.}
\label{fig:peakba}
\vspace{-2ex}
\end{figure}

\paragraph{\bf Effectiveness and persistence.}
Continuous attacks achieve considerably higher peak BA than sparse attacks --- mean peak values of 77.1\% and 72.6\% under FedAvg and FedProx, respectively, compared to 56.7\% and 54.2\% for sparse attacks (Figure~\ref{fig:peakba}; Table~\ref{tab:summary}). Peak effectiveness, however, is a poor predictor of long-term threat. Under FedAvg at the lowest gradient mask ratio, sparse attacks require 250 post-attack rounds to stabilize below the 25\% BA threshold, while continuous attacks stabilize below it in 53 rounds, indicating different decay dynamics: continuous poisoning creates a strong but fast-eroding backdoor, whereas sparse gradual injection embeds a more dilution-resistant one.

\paragraph{\bf Durability.}
Both  Method~1 (threshold-based durability) and Method~2 (volatility-based durability) show that backdoor persistence is non-linear and threshold-dependent. Under Method~1, backdoor accuracy decays rapidly above 40\% but persists for hundreds of rounds at lower levels (as shown in Table~\ref{tab:durability}); the most persistent configuration --- a sparse attack at mask ratio 0.05 under FedAvg --- never stabilizes below 20\% within 919 post-attack rounds. Method~2 reveals that the volatility of backdoor follows a characteristic two-stage pattern. Large fluctuations subside rapidly, with the rolling standard deviation falling within $\pm$10\% in roughly 20--50 rounds across most configurations. 
Convergence to finer stability is far slower: reaching $\pm$5\% requires several hundred post-attack rounds, and $\pm$3\% is slower still under FedProx, several configurations never attain this level within the experiment. Notably, the overall post-attack standard deviation remains within a tight band even as attack pattern and gradient mask ratio vary widely, the same parameters that strongly affect peak BA and persistence. In contrast, the band shifts noticeably between FedAvg (5.8--6.4\%) and FedProx (6.8--7.4\%). 
This shows that residual backdoor volatility depends on aggregation choice, not attack configuration. Behavioral stability and elimination are therefore distinct: a backdoor may stabilize yet remain active at non-trivial accuracy. These findings show that honest aggregation alone cannot remove embedded backdoors and that sparse, context-triggered attacks are especially stealthy and durable threats to federated surveillance systems.

\begin{table}[t]
\centering
\caption{Stealth and peak backdoor accuracy summary (mean across gradient mask ratios)}
\label{tab:summary}
\begin{tabular}{llcc}
\toprule
Pattern & Aggregation & Clean Acc.\ Drop & Peak BA \\
\midrule
Sparse     & FedAvg  & 1.66\% & 56.7\% \\
Continuous & FedAvg  & 7.51\% & 77.1\% \\
Sparse     & FedProx & 1.03\% & 54.2\% \\
Continuous & FedProx & 4.23\% & 72.6\% \\
\bottomrule
\end{tabular}
\end{table}

\begin{table}[t]
\centering
\caption{Durability summary: mean rounds to stabilise below the 25\% threshold (Method~1) and mean post-attack standard deviation (Method~2), averaged across gradient mask ratios. Full per-ratio results in Appendix~\ref{app:method12}}
\label{tab:durability}
\vspace{-1.5ex}
\begin{tabular}{llcc}
\toprule
Pattern & Aggregation & Stabilise $<$25\% & Post-attack $\sigma$ \\
\midrule
Sparse     & FedAvg  & 120 & 6.1\% \\
Continuous & FedAvg  & 82  & 6.3\% \\
Sparse     & FedProx & 119 & 7.0\% \\
Continuous & FedProx & 136 & 7.1\% \\
\bottomrule
\end{tabular}
\end{table}

\section{Conclusion and Future Work}
We presented STAIN-FL, a contextual backdoor injection framework for federated video anomaly detection. Instead of using artificial trigger patterns, STAIN-FL exploits natural surveillance conditions and combines anomaly-to-benign relabeling with masked malicious updates to improve stealth and persistence. Experimental results show that sparse attacks are the most concerning setting, keeping clean-accuracy drops low, at $1.66\%$ under FedAvg and $1.03\%$ under FedProx, while still reaching peak backdoor accuracies of $56.7\%$ and $54.2\%,$ respectively. Under FedAvg, sparse attacks remain above the $25\%$ backdoor-accuracy threshold for an average of $336$ post-attack rounds. FedProx reduces some attack effects, but does not eliminate long-term persistence. These findings show that continued benign aggregation alone is insufficient to remove embedded backdoors. Future work will study agentic backdoor attacks, in which the compromised client adaptively selects triggers, timing, mask ratios, and poisoning strength based on FL dynamics.

\bibliographystyle{ACM-Reference-Format}
\bibliography{flad}

\appendix

\section{Per-Client Data Distribution}\label{app:data}
The non-IID split assigns each client domain-specific anomaly classes, with selected classes shared to reflect overlapping agency boundaries. The server holds a balanced global test set, while the remaining data is distributed across four clients and split locally into 80/20 train/test sets. Table~\ref{tab:perclient} shows the resulting distribution.

\begin{table}[!htb]
\centering
\caption{Non-IID federated data distribution. The server holds the global test set, while client data is split across four agencies, showing Normal and Anomaly counts}
\label{tab:perclient}
\small
\begin{tabular}{lccc}
\toprule
Partition & Normal & Anomaly & Total \\
\midrule
Global test set (server) & 190 & 190 & 380 \\
\midrule
Client~1 & 282 & 300 & 582 \\
Client~2 & 221 & 220 & 441 \\
Client~3 & 189 & 160 & 349 \\
Client~4 & 68  & 80  & 148 \\
\midrule
Client total & 760 & 760 & 1{,}520 \\
\midrule
\textbf{Overall} & \textbf{950} & \textbf{950} & \textbf{1{,}900} \\
\bottomrule
\end{tabular}
\vspace{-2ex}
\end{table}

\section{Stealth Threshold Justification}\label{app:threshold}
The $5\%$ stealth threshold is chosen to exceed the natural $\pm$2--3\% per-round accuracy fluctuation observed under benign non-IID FL training. This margin avoids flagging normal variance as an attack, while still capturing sustained performance degradation. From a defender's perspective, drops below $5\%$ are difficult to distinguish from routine training noise, whereas larger drops are more likely to indicate abnormal behavior.

\section{Impact Rounds Analysis}\label{app:impact}
As a complementary durability measure, the impact rounds analysis counts the total number of post-attack rounds in which backdoor accuracy remained above a given threshold. Unlike Method~1, which records the round at which the backdoor permanently stabilises below a threshold, this measure tallies the cumulative number of effective rounds. Table~\ref{tab:impact} reports these counts for all configurations.

\begin{table}[!htb]
\centering
\caption{Impact rounds analysis on the number of post-attack rounds with backdoor accuracy above each threshold}
\label{tab:impact}
\small
\setlength{\tabcolsep}{4pt}
\begin{tabular}{llcccccccc}
\toprule
 & & \multicolumn{2}{c}{$\geq$50\%} & \multicolumn{2}{c}{$\geq$40\%} & \multicolumn{2}{c}{$\geq$30\%} & \multicolumn{2}{c}{$\geq$25\%} \\
\cmidrule(lr){3-4}\cmidrule(lr){5-6}\cmidrule(lr){7-8}\cmidrule(lr){9-10}
Pattern & $k$ & FA & FP & FA & FP & FA & FP & FA & FP \\
\midrule
\multirow{5}{*}{Sparse}
 & 0.05 & 3 & 2 & 16 & 22 & 169 & 168 & 382 & 358 \\
 & 0.10 & 3 & 2 & 19 & 18 & 162 & 156 & 387 & 327 \\
 & 0.15 & 2 & 1 & 13 & 18 & 126 & 138 & 312 & 305 \\
 & 0.20 & 2 & 3 & 13 & 24 & 110 & 128 & 308 & 275 \\
 & 0.25 & 3 & 2 & 13 & 20 & 116 & 135 & 293 & 273 \\
\midrule
\multirow{5}{*}{Continuous}
 & 0.05 & 5  & 4 & 19 & 21 & 117 & 152 & 290 & 347 \\
 & 0.10 & 7  & 6 & 15 & 25 & 121 & 136 & 299 & 272 \\
 & 0.15 & 8  & 6 & 16 & 27 & 113 & 151 & 307 & 320 \\
 & 0.20 & 10 & 6 & 18 & 27 & 126 & 170 & 312 & 359 \\
 & 0.25 & 8  & 7 & 15 & 24 & 126 & 136 & 297 & 280 \\
\bottomrule
\end{tabular}
\end{table}

\begin{table}[t]
\centering
\caption{Threshold-based stabilization (Method~1): rounds for backdoor accuracy to fall and remain below each threshold. FA: FedAvg; FP: FedProx}
\label{tab:method1full}
\small
\setlength{\tabcolsep}{4pt}
\begin{tabular}{llcccccccccc}
\toprule
 & & \multicolumn{2}{c}{50\%} & \multicolumn{2}{c}{40\%} & \multicolumn{2}{c}{30\%} & \multicolumn{2}{c}{25\%} & \multicolumn{2}{c}{20\%} \\
\cmidrule(lr){3-4}\cmidrule(lr){5-6}\cmidrule(lr){7-8}\cmidrule(lr){9-10}\cmidrule(lr){11-12}
Pattern & $k$ & FA & FP & FA & FP & FA & FP & FA & FP & FA & FP \\
\midrule
\multirow{5}{*}{Sparse}
 & 0.05 & 20 & 20 & 20 & 20 & 40 & 35 & 250 & 110 & NA  & 506 \\
 & 0.10 & 20 & 20 & 20 & 20 & 35 & 63 & 127 & 134 & 531 & 647 \\
 & 0.15 & 20 & 20 & 20 & 23 & 24 & 36 & 58  & 84  & 88  & 602 \\
 & 0.20 & 20 & 20 & 20 & 21 & 23 & 37 & 55  & 160 & 77  & 462 \\
 & 0.25 & 20 & 20 & 21 & 20 & 42 & 34 & 111 & 109 & 240 & 510 \\
\midrule
\multirow{5}{*}{Continuous}
 & 0.05 & 20 & 20 & 22 & 20 & 26 & 49 & 53  & 198 & NA  & 738 \\
 & 0.10 & 20 & 20 & 22 & 21 & 37 & 29 & 81  & 90  & 421 & 412 \\
 & 0.15 & 20 & 20 & 20 & 20 & 23 & 37 & 70  & 92  & 129 & 504 \\
 & 0.20 & 20 & 20 & 20 & 20 & 25 & 40 & 95  & 179 & 326 & 379 \\
 & 0.25 & 20 & 20 & 20 & 20 & 33 & 43 & 110 & 120 & 466 & 315 \\
\bottomrule
\end{tabular}
\vspace{-2ex}
\end{table}

\begin{table}[t]
\centering
\caption{Method~2 volatility stabilization: rounds until backdoor-accuracy variation stays within each tolerance. FA: FedAvg; FP: FedProx}
\label{tab:method2full}
\small
\setlength{\tabcolsep}{4pt}
\begin{tabular}{llcccccccc}
\toprule
 & & \multicolumn{2}{c}{$\pm$10\%} & \multicolumn{2}{c}{$\pm$5\%} & \multicolumn{2}{c}{$\pm$3\%} & \multicolumn{2}{c}{Post-attack $\sigma$} \\
\cmidrule(lr){3-4}\cmidrule(lr){5-6}\cmidrule(lr){7-8}\cmidrule(lr){9-10}
Pattern & $k$ & FA & FP & FA & FP & FA & FP & FA & FP \\
\midrule
\multirow{5}{*}{Sparse}
 & 0.05 & 44 & 47 & 353 & 368 & 886 & NA  & 6.3\% & 7.0\% \\
 & 0.10 & 24 & 33 & 377 & 512 & 806 & 574 & 6.3\% & 7.2\% \\
 & 0.15 & 29 & 37 & 360 & 301 & 754 & 359 & 6.2\% & 6.8\% \\
 & 0.20 & 21 & 40 & 279 & 380 & 708 & 612 & 5.8\% & 6.9\% \\
 & 0.25 & 48 & 40 & 320 & 379 & 478 & 786 & 6.1\% & 6.9\% \\
\midrule
\multirow{5}{*}{Continuous}
 & 0.05 & 40 & 44 & 303 & 352 & 637 & NA  & 6.3\% & 7.0\% \\
 & 0.10 & 33 & 77 & 408 & 434 & 870 & 849 & 6.3\% & 7.4\% \\
 & 0.15 & 34 & 64 & 346 & 398 & 850 & NA  & 6.1\% & 7.2\% \\
 & 0.20 & 44 & 54 & 360 & 338 & 883 & 771 & 6.4\% & 7.1\% \\
 & 0.25 & 41 & 56 & 446 & 328 & 674 & NA  & 6.2\% & 7.0\% \\
\bottomrule
\end{tabular}
\vspace{1.5ex}
\end{table}

\section{Full Stabilization Results}
\label{app:method12}

Tables~\ref{tab:method1full} and~\ref{tab:method2full} report the full post-attack stabilization results. Table~\ref{tab:method1full} presents the threshold-based stabilization rounds for all gradient mask ratios, aggregation algorithms, and accuracy thresholds. A value of $20$ indicates immediate stabilization within the first measurement window, while NA indicates that the backdoor did not stabilize below the stated threshold during the post-attack period.
Table~\ref{tab:method2full} presents the volatility-based stabilization rounds for all configurations across the three volatility tolerances, together with the overall post-attack standard deviation. NA indicates that the stated volatility tolerance was not reached in the post-attack period.

\section{Global Model Performance across FL Rounds}
\label{app:graphs}

This section reports the global main and backdoor accuracy across all twenty configurations. Each plot shows the per-round mean over five runs, clipped to the first 500 rounds for readability. Figures~\ref{fig:app-fedavg-k005}--\ref{fig:app-fedprox-k025} are grouped by aggregation algorithm and mask ratio $k$, with one panel per attack pattern and a shared legend as shown below.
\begin{center}
\includegraphics[width=0.47\textwidth]{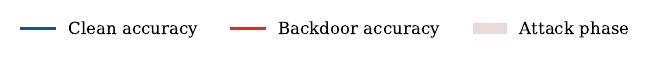}
\end{center}

\newcommand{\appFigW}{0.48\textwidth}
\newcommand{\appPanelH}{2.6cm}

\newcommand{\appFig}[4]{%
  \begin{minipage}[h]{\appFigW}
    \centering
    \begin{minipage}{0.49\linewidth}
      \centering
      \includegraphics[height=\appPanelH]{#1}\\[2pt]
      {\small (a) Sparse}
    \end{minipage}\hfill
    \begin{minipage}{0.49\linewidth}
      \centering
      \includegraphics[height=\appPanelH]{#2}\\[2pt]
      {\small (b) Continuous}
    \end{minipage}\\[4pt]
    \captionof{figure}{#3}
    \label{#4}
  \end{minipage}%
}

\begin{figure*}[h]
\centering
\appFig{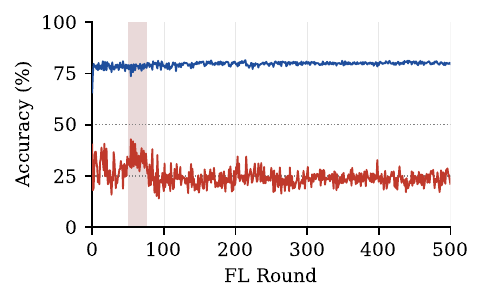}{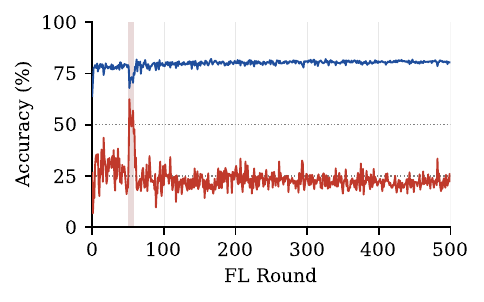}%
       {FedAvg, $k=0.05$: accuracy over FL rounds.}{fig:app-fedavg-k005}
\hfill
\appFig{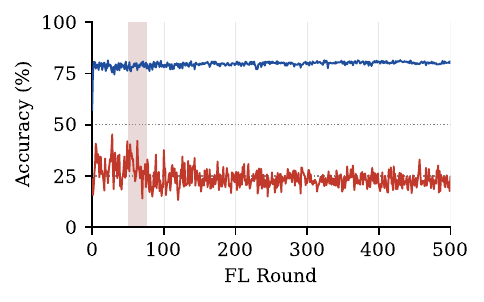}{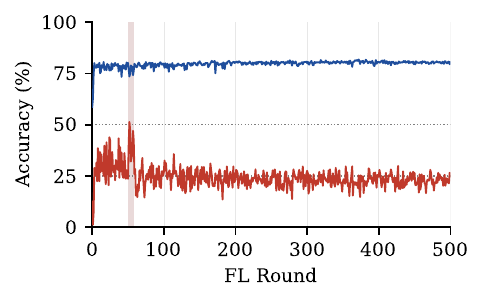}%
       {FedProx, $k=0.05$: accuracy over FL rounds.}{fig:app-fedprox-k005}
\end{figure*}

\begin{figure*}[h]
\centering
\appFig{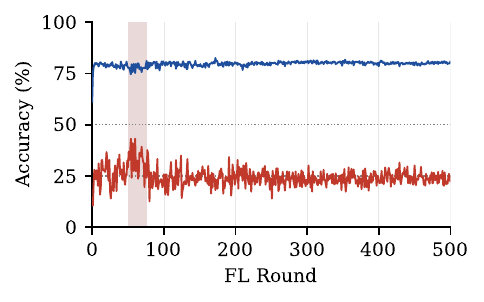}{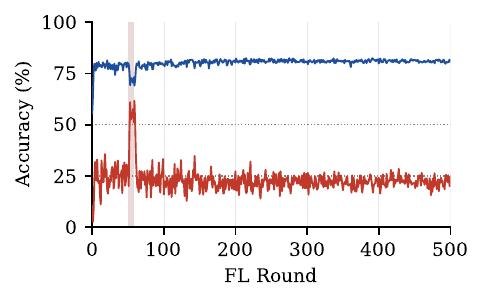}%
       {FedAvg, $k=0.10$: accuracy over FL rounds.}{fig:app-fedavg-k010}
\hfill
\appFig{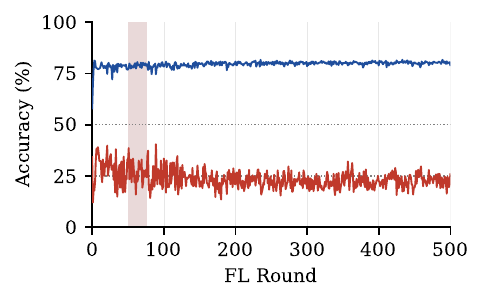}{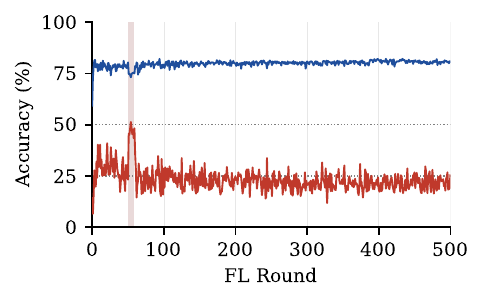}%
       {FedProx, $k=0.10$: accuracy over FL rounds.}{fig:app-fedprox-k010}

\vspace{14pt}

\appFig{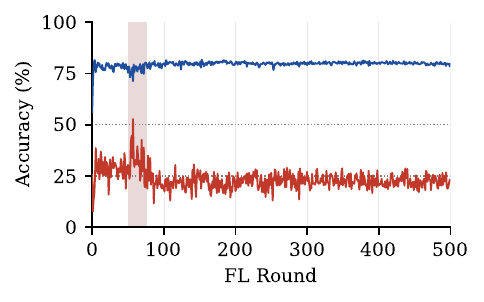}{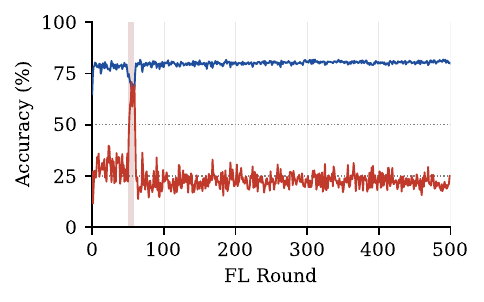}%
       {FedAvg, $k=0.15$: accuracy over FL rounds.}{fig:app-fedavg-k015}
\hfill
\appFig{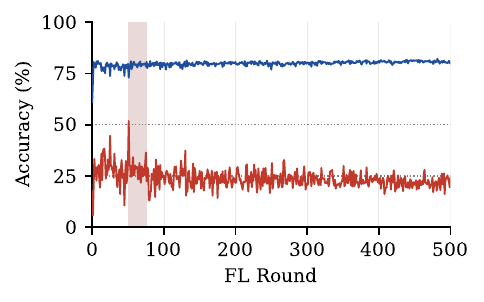}{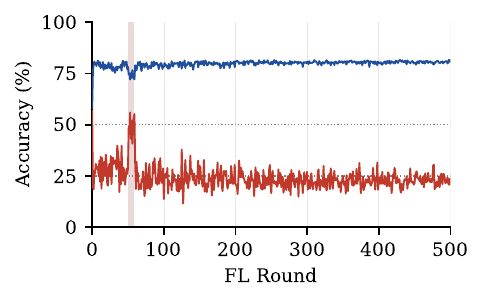}%
       {FedProx, $k=0.15$: accuracy over FL rounds.}{fig:app-fedprox-k015}

\vspace{14pt}

\appFig{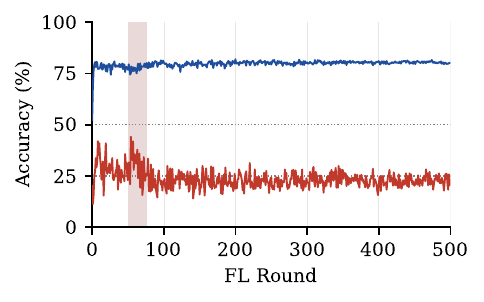}{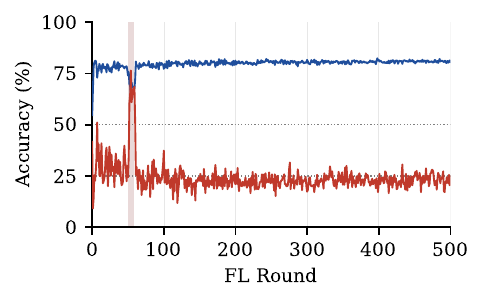}%
       {FedAvg, $k=0.20$: accuracy over FL rounds.}{fig:app-fedavg-k020}
\hfill
\appFig{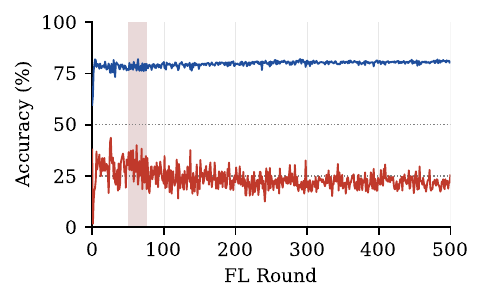}{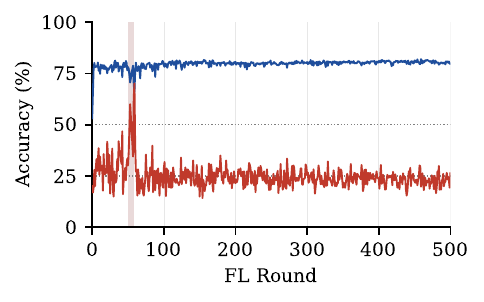}%
       {FedProx, $k=0.20$: accuracy over FL rounds.}{fig:app-fedprox-k020}
\end{figure*}

\begin{figure*}[h]
\centering
\appFig{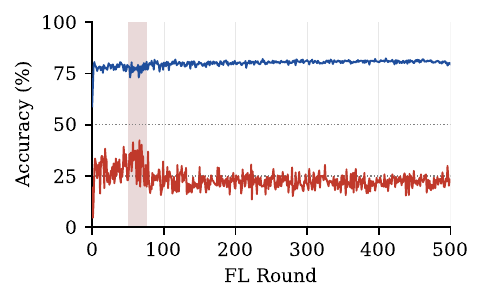}{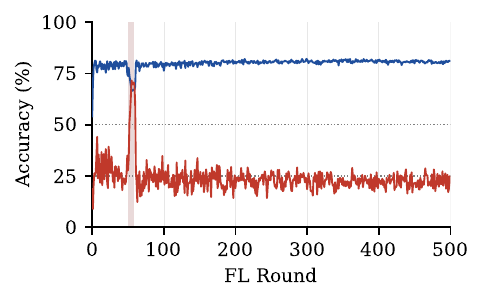}%
       {FedAvg, $k=0.25$: accuracy over FL rounds.}{fig:app-fedavg-k025}
\hfill
\appFig{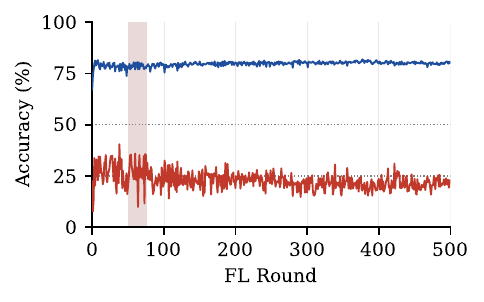}{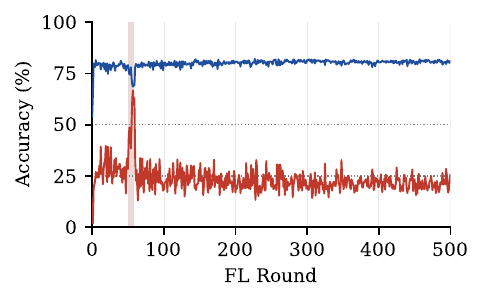}%
       {FedProx, $k=0.25$: accuracy over FL rounds.}{fig:app-fedprox-k025}
\end{figure*}

\end{document}